\documentclass[authorversion=true,sigconf]{acmart}

\AtBeginDocument{%
  \providecommand\BibTeX{{%
    \normalfont B\kern-0.5em{\scshape i\kern-0.25em b}\kern-0.8em\TeX}}}

\setcopyright{acmcopyright}
\copyrightyear{2022}
\acmYear{2022}
\setcopyright{rightsretained}
\acmConference[ESEM '22]{ACM / IEEE International Symposium on Empirical Software Engineering and Measurement (ESEM)}{September 19--23, 2022}{Helsinki, Finland}
\acmBooktitle{ACM / IEEE International Symposium on Empirical Software Engineering and Measurement (ESEM) (ESEM '22), September 19--23, 2022, Helsinki, Finland}\acmDOI{10.1145/3544902.3546252}
\acmISBN{978-1-4503-9427-7/22/09}

\usepackage{tabularx}
\usepackage{nicefrac}
\usepackage{graphicx}
\usepackage{dblfloatfix}
\usepackage{fancybox}

\usepackage{xcolor}
\def\BibTeX{{\rm B\kern-.05em{\sc i\kern-.025em b}\kern-.08em
		T\kern-.1667em\lower.7ex\hbox{E}\kern-.125emX}}
\usepackage{framed}
\usepackage{tcolorbox}
\tcbuselibrary{theorems}

\newtcbtheorem[number within=section]{definition}{Definition}%
{colback=gray!5,colframe=gray!10!gray,fonttitle=\bfseries}{th}
\usepackage[strict]{changepage}
\definecolor{formalshade}{gray}{0.96}
\definecolor{darkblue}{RGB}{0,80,155}
\definecolor{green}{RGB}{177,201,31}

\begin{document}

\title[Meetings and Mood -- Related or Not? Insights from Student Software Projects]{Meetings and Mood -- Related or Not?\\Insights from Student Software Projects}


\author{Jil Klünder}
\email{jil.kluender@inf.uni-hannover.de}
\orcid{1234-5678-9012}
\affiliation{%
	\institution{Leibniz University Hannover\\Software Engineering Group}
	\streetaddress{Welfengarten 1}
	\city{Hannover}
	\state{}
	\country{Germany}
	\postcode{30167}
}

\author{Oliver Karras}
\email{oliver.karras@tib.eu}
\affiliation{%
	\institution{TIB - Leibniz Information Centre\\for Science and Technology}
	\streetaddress{Welfengarten 1B}
	\city{Hannover}
	\state{}
	\country{Germany}
	\postcode{30167}
}

\renewcommand{\shortauthors}{Klünder and Karras}

\begin{abstract}
	[Background:] Teamwork, coordination, and communication are a prerequisite for the timely completion of a software project. Meetings as a facilitator for coordination and communication are an established medium for information exchange. Analyses of meetings in software projects have shown that certain interactions in these meetings, such as proactive statements followed by supportive ones, influence the mood and motivation of a team, which in turn affects its productivity. So far, however, research has focused only on certain interactions at a detailed level, requiring a complex and fine-grained analysis of a meeting itself. 
	
	\noindent
	[Aim:] In this paper, we investigate meetings from a more abstract perspective, focusing on the polarity of the statements, i.e., whether they appear to be positive, negative, or neutral. 
	
	\noindent
	[Method:] We analyze the relationship between the polarity of statements in meetings and different social aspects, including conflicts as well as the mood before and after a meeting. 
	
	\noindent
	[Results:] Our results emerge from 21 student software project meetings and show some interesting insights: (1) Positive mood before a meeting is both related to the amount of positive statements in the beginning, as well as throughout the whole meeting, (2) negative mood before the meeting only influences the amount of negative statements in the first quarter of the meeting, but not the whole meeting, and (3) the amount of positive and negative statements during the meeting has no influence on the mood afterwards. 
	
	\noindent
	[Conclusions:] We conclude that the behaviour in meetings might rather influence short-term emotional states (feelings) than long-term emotional states (mood), which are more important for the project.
\end{abstract}

\ccsdesc[500]{Software and its engineering~Programming teams}
\ccsdesc[500]{Software and its engineering~Collaboration in software development}

\keywords{Software development teams, sentiment analysis, meeting, mood}

\maketitle

\section{Introduction}
Meetings are an established medium to share information in software projects~\cite{ambler2002agile,schneider2018positive,kluender2017team}. If they are well organized and if the participants interact adequately, meetings are an efficient way to transport a lot of information in a short amount of time~\cite{oshri2007global}. However, meetings are not always as effective as they might be, causing demotivation and frustration~\cite{kauffeld2012meetings,klunder2020you}. Besides the effectivity of a meeting, other factors such as interactions and how the participants communicate play a major role~\cite{cohen2011meeting,luong2005meetings,rogelberg2006not}. Focusing on software development teams, Schneider et al.~\cite{schneider2018positive} showed that proactive statements followed by supportive statements in a meeting increase positive affect afterwards.

Schneider et al.~\cite{schneider2018positive} applied so-called interaction analysis in meetings by focusing on interactions such as interrupting other participants, blaming, and praising~\cite{kauffeld2012meetings,prenner2018making,klunder2020you}. Interactions, including statements, are assigned to a code according to their meaning, e.g., whining, describing a problem, connecting problem and solution, and getting lost in details~\cite{klunder2020you,kauffeld2012meetings}. Most interactions can be advantageous or disadvantageous for meetings (depending on the intention and the situation)~\cite{kauffeld2012meetings}. For example, talking about problems is often perceived as not-negative, as only known problems can be solved~\cite{kauffeld2009complaint}. Given this fact, interaction analysis provides an objective picture of the meeting, without allowing conclusions on whether the team is satisfied with the meeting or not.

So far, the influence of meetings on social aspects, such as mood, motivation, and productivity, has only been partially explored \cite{klunder2020affecting}. For example, it is unclear whether and how a meeting influences the mood of a team afterwards. However, given the amount of time most software project team members spend in meetings, knowing about the influence of meetings on the project (including team-related aspects such as motivation and productivity) is of particular importance~\cite{klunder2020you,rogelberg2006not,luong2005meetings}. 

In this paper, we analyze the statements made in meetings according to their polarity (i.e., whether they are positive, negative, or neutral) regarding their relation to social aspects. We analyze 21 student software projects to investigate (1) whether there is an influence of the mood of a team before the meeting on the polarity of the statements made during the meeting, (2) whether the polarity of the statements has an influence on the mood of a team after the meeting, and (3) whether the polarity of the statements is related to the perceived risk of social or task-related conflicts during the project. In these projects, the first team meeting was recorded on video and the participants reported on various social aspects.

Despite the small sample size, we make three noteworthy observations: (1) Students intuitively behave adequately in meetings, having way more positive than negative statements during the meetings, (2) starting a meeting with a high positive mood can smooth both the meeting start as well as the meeting as a whole, and (3) the polarity of statements made during the meeting has no measurable influence on the mood afterwards. 

\textit{Outline.} The rest of the paper is structured as follows: In Section~\ref{sec:background-and-related-work}, we present background information and related work. Section~\ref{sec:research-design} summarizes our research design with the research questions, the data base, the data collection, and the data analysis procedures. In Section~\ref{sec:results}, we present the results which we discuss in Section~\ref{sec:discussion}. Section~\ref{sec:conclusion} concludes the paper and presents future work.  

\section{Background and Related Work}\label{sec:background-and-related-work}
Meeting analysis has frequently been subject to software engineering research. Based on insights from 20 student software projects, Schneider et al.~\cite{schneider2015media} investigate relationships between media, mood, and meetings. Liskin et al.~\cite{liskin2013meeting} analyze meeting profiles of 14 student development teams based on the duration and the frequency in different project phases. Their results show a relation between the meeting behavior and perceived pressure during the project. 

In 1992, Olson et al. \cite{olson1992small} investigated interactions in meetings of development teams. They analyzed ten design meetings in four projects by assigning each statement made during the meeting to a category such as ``issue'', ``project management'' or ``meeting management''. Schneider et al.~\cite{schneider2018positive} analyzed the behavior of 155 student software developers in 32 teams during their first team meeting. They used an established coding scheme from psychology. Their results show a significant positive influence of proactive statements on a team's affect. Developers tend to be more satisfied if proactive behavior was present during the meeting. This effect increased if proactive statements were followed by supportive statements. Prenner et al.~\cite{prenner2018making} analyzed the relation between meeting satisfaction and the occurrence of single categories used by Schneider et al.~\cite{schneider2018positive}. 

Besides the meeting analysis on interaction level, some authors analyzed the content of a meeting. Gall and Berenbach~\cite{gall2006towards} present a framework that automatically extracts information in elicitation meetings. Shakeri et al.~\cite{shakeri2018elica} also support elicitation meetings by automatically extracting knowledge related to requirements. In contrast to these two approaches, Karras et al. \cite{karras2016supporting, karras2021supporting} developed a manual approach that combines textual minutes and video recordings to analyze the content of elicitation meetings.

Herrmann and Klünder~\cite{herrmann2021textual} propose to apply sentiment analysis to statements in meetings. This approach provides an overview of the polarity of the statements in a meeting, i.e., whether the majority of statements is positive, negative, or neutral. Their concept processes live meeting audio data which is classified into sentiment polarity classes. They tested their concept in a preliminary study with a student software project team in comparison to the results of a human observer, showing moderate agreement~\cite{herrmann2021textual}. 

Sentiment analysis has frequently been applied to text-based communication in software projects~\cite{obaidi2021development}. According to a recent literature review~\cite{obaidi2021development}, there are three main topics in software engineering research on sentiment analysis: (1) Development of sentiment analysis tools, (2) comparison of different tools, and (3) application of the tools in specific contexts. As in this paper sentiment analysis is used as an analysis method, we are mainly interested in existing tools that fit best for our matter. For the software engineering domain, there are three tools: SentiStrength-SE, Senti4SD, and SentiCR~\cite{obaidi2021development}. These tools are trained using different data bases, leading to a different context of use. For example, SentiCR is designed for code reviews. As our analysis is based on a German data set, we need a German sentiment analysis tool. However, as far as we know, there is only one software engineering specific attempt to train a classifier based on German data~\cite{klunder2020identifying}. Klünder et al.~\cite{klunder2020identifying} use communication data from Zulip, the communication tool used in that software project. The authors label the data themselves and use it as input to train the classifier. They reach an accuracy of 62.97\%, which is mainly due to the small size of the data set.

The research presented in this paper uses sentiment analysis to draw conclusions between social aspects and the polarity of statements in meetings. That is, we use sentiment analysis as a method without making it subject to our research. This way, we investigate if the mood in a team is related to meeting behavior, and if meeting behavior is related to the mood and conflicts afterwards.

\section{Research Design}\label{sec:research-design}
In the following, we present the research design with the underlying research questions, information about the used data set, and the data analysis procedures.

\subsection{Research Goal and Research Questions}\label{sec:research-goal-and-research-questions}
The overall goal of the research presented in this paper is to investigate relationships between the mood of a team and the statements made in a meeting. In particular, we wanted to analyze relations between (1) the mood before a meeting, (2) the polarity of statements in the meeting, and (3) the mood and other social aspects after a meeting. For this purpose, we ask the following research questions:\\

\noindent
\textbf{RQ1:} \textit{How are the mood of a team before a meeting, the polarity of statements made during the meeting, and the team mood after a meeting related to each other?}\vspace{3mm}

\noindent
This question asks for a relationship between mood and meetings. In particular, we analyze whether teams in a good or bad mood make more positive or negative statements during the meetings, and whether an increased number of positive or negative statements in a meeting is related to a better or worse mood afterwards.\\

\noindent
\textbf{RQ2:} \textit{How is the polarity of the statements made during a meeting related to the likelihood of social or task-related conflicts in the team?}\vspace{3mm}

\noindent
Besides the mood, there are several other aspects that influence the collaboration in a team. This research question asks for a relation between the polarity of the statements made in a meeting and whether conflicts are likely to be present throughout the project.\\

\noindent 
\textbf{RQ3:} \textit{How do the relationships between polarity and mood change when only looking at meeting start and end?}\vspace{3mm}

\noindent
To provide more detailed insights, we also want to analyze whether the influences affect the meeting as a whole or just the first minutes (respectively that the last minutes mainly influence the mood afterwards). 

\subsection{Data Set}
Our analysis is based on two data sets used by Schneider et al.~\cite{schneider2018positive, schneider2015media}. The data sets emerged from student software projects conducted at Leibniz University Hannover.

\subsubsection{Data Source: Student Software Projects}
The Software Engineering Group at Leibniz University Hannover offers a yearly course for computer science students who are close to achieving their bachelor degree. Usually, students in their third year (fifth semester) of their bachelor studies participate in the course. In teams consisting of three to five members, the students experience all phases of a software project, including requirements elicitation, implementation, testing, acceptance tests, and the delivery. Over the project duration of 15 weeks, the students need to organize themselves, coordinate their communication and the communication with the customer.

In these 15 weeks, the students develop a software for a customer who is in most cases part of the Software Engineering Group. Exemplary projects are the development of a card game, the visualization of communication, and a game with a labyrinth. Each student is expected to invest about 270 hours in their project. 

\subsubsection{Data Collection} 
In the winter semesters starting in 2012 and 2013, the organizers of the course collected different kind of data to be used to understand social aspects in (student) development teams~\cite{schneider2015media,schneider2018positive}. As, to the best of our knowledge, this is one of the largest data sets consisting of different social aspects (including reportings on conflicts, affect, communication) with recordings of meetings, we decided to use this data set for our study. 

In these two years, in total 165 students participated in the software project, organized in 34 arbitrarily formed teams. The first team meeting right after the first meeting with the customer was video-recorded and transcribed \cite{schneider2018positive}. In addition, throughout the project, the teams were asked to report on various psychological aspects, their meeting and communication behavior~\cite{schneider2015media}. The psychological aspects included positive and negative affect, satisfaction, task-related and social conflicts. This data was collected using a paper-pencil-questionnaire. While positive and negative affect were rated before and after the meeting, all other variables were only collected after the meeting (and weekly thereafter). For data collection, established scales from psychology were used, e.g., the positive and negative affect schedule (PANAS)~\cite{watson1988development} in its German version~\cite{krohne1996untersuchungen}. The data is organized in two data sets: The first data set contains all of the psychological information collected over the full project duration. The second data set contains meeting transcripts from 32 teams, although there were problems with the recordings in two cases. The meetings were manually transcribed by PhD students from psychology and software engineering.

\subsubsection{Ethics Vote}
The ethics committee at Leibniz Universität Hannover authorized this collection of data. The students were informed about the data collection and the further usage. In particular, the students got the guarantee that the data will not be published at any time (neither the recordings or the transcripts, nor the other data). Thus, we are not allowed to make the data publicly available. Nevertheless, all data records were anonymized and the data did not influence the success of passing the course.  

\subsubsection{Variable Selection}
For the analyses presented in this paper, we did not use the full data set. Instead, we concentrated on the following data related to our research questions: 
\begin{itemize}
	\item \textit{Meeting transcripts} containing textual representations of the statements made during the meeting. The meetings were manually transcribed by psychology and software engineering researchers involved in the project~\cite{schneider2018positive}.\vspace{3mm} 
	\item \textit{Positive and negative affect before and after the meeting} collected using the established PANAS scale~\cite{watson1988development} in its German version~\cite{krohne1996untersuchungen}. For both positive and negative affect, ten items that are rated on a 5-point Likert scale (from never to always) by each team member. Table~\ref{tab:panas} summarizes the items for positive and negative affect. We describe the aggregation of these items to team mood in Table~\ref{tab:variables}.\vspace{3mm}
	\item 
	\textit{Social and task-related conflicts after the meeting} collected using Jehn's intragroup conflict scale~\cite{jehn1995multimethod} in its German version~\cite{lehmann2011task}. For both types of conflict, four items that are rated on a 6-point Likert scale (from never / none to very often / very much). The four items for each of the scales are summarized in Table~\ref{tab:conflicts}. We describe the aggregation of these items to team mood in Table~\ref{tab:variables}.
\end{itemize}

\begin{table}[!htbp]
	\small
	\caption{Items related to positive and negative affect~\cite{watson1988development}}
	\label{tab:panas}
	\begin{tabularx}{\columnwidth}{p{0.45\columnwidth}p{0.45\columnwidth}}
		\toprule 
		Positive Affect & Negative Affect\\
		\midrule
			active &afraid\\
			attentive &ashamed\\
			alert &distressed \\
			excited &guilty\\
			enthusiastic &hostile \\
			determined &irritable \\
			inspired &jittery \\
			interested &nervous \\
 			proud &scared \\
			strong &upset \\
		\bottomrule 
	\end{tabularx} 
\end{table}

\begin{table}[!h]
	\small
	\caption{Items related to relationship and task-related conflicts~\cite{jehn1995multimethod}}
	\label{tab:conflicts}
	\begin{tabularx}{\columnwidth}{p{0.45\columnwidth}p{0.45\columnwidth}}
		\toprule 
		Relationship conflicts & Task-related conflicts\\
		\midrule
		(1) How much friction is there among members in your work unit? &(1) How often do people in
		your work unit disagree about opinions regarding the work being done?\\
		(2) How much are personality conflicts evident in your work unit? &(2) How frequently are there conflicts about ideas in your work unit?\\
		(3) How much tension is there among members in your work unit? &(3) How much conflict about the work you do is there in your work unit?  \\
		(4) How much emotional conflict is there among members in your work unit? &(4) To what extent are there
		differences of opinion in your work unit? \\
		\bottomrule 
	\end{tabularx} 
\end{table}

\subsubsection{Data Selection and Cleaning}\label{sec:data-selection-and-cleaning}
As part of the data cleaning process, we removed transcripts that were obviously incomplete in whole or in part, e.g., only ten statements made in a one-hour meeting. These incomplete transcripts emerged from a low audio quality. We removed data from eleven teams, leading to a total of 21 teams included in our study.

As part of the quality assurance, we went through the remaining 21 transcripts and removed all entries that represent ``not applicable'' values (marked by ``unv.'', which means ``incomprehensible''). Note that removing single entries did not lead to an exclusion of the whole team from further analysis, as we assume that the polarity of a single statement in a meeting does not impact the entire meeting.

In the end, we considered the data of 21 teams with a total of 102 participants as suitable for our analysis. Most of the teams (19) consisted of 5 persons, with one team having 3 and another one having four team members. The students had an average age of 23 years, and most of them (94 out of 102) reported being male.
The meetings had an average duration of 39 minutes and 49 seconds (min: 7 minutes and 52 seconds, max: 74 minutes and 5 seconds, SD: 18 minutes and 3 seconds), and on average, 742 statements were made in the meeting (min: 162, max: 1858, SD: 3975).

\begin{table*}[!htbp]
	\small
	\caption{Used variables, null hypotheses on the relations between the variables, \\and their relation to the research questions}
	\label{tab:HypothesisOverview}
	\begin{tabularx}{\textwidth}{p{0.06\columnwidth}Xp{0.06\columnwidth}}
		\toprule 
		\multicolumn{2}{l}{\textbf{Hypotheses}} & \textbf{RQ}\\
		\midrule
		H1$_0$ & \textbf{There is no difference between the amount of positive and negative statements in the meeting.}&RQ1\\		
		\midrule
		H2$_0$ & \textbf{The polarity of the statements in the meeting does not depend on the mood before the meeting.}&RQ1\\
		H2.1$_{0}$ & The amount of positive statements does not depend on the positive mood before the meeting. &RQ1\\ 
		H2.2$_{0}$ & The amount of positive statements does not depend on the negative mood before the meeting.&RQ1 \\
		H2.3$_{0}$ & The amount of negative statements does not depend on the positive mood before the meeting.&RQ1 \\ 
		H2.4$_{0}$ & The amount of negative statements does not depend on the negative mode before the meeting.&RQ1 \\ 
		\midrule
		H3$_0$ & \textbf{The mood after the meeting does not depend on the polarity of the statements during the meeting.}&RQ1\\
		H3.1$_{0}$ & The positive mood after the meeting does not depend on the amount of positive statements during the meeting.&RQ1 \\ 
		H3.2$_{0}$ & The positive mood after the meeting does not depend on the amount of negative statements during the meeting.&RQ1 \\
		H3.3$_{0}$ & The negative mood after the meeting does not depend on the amount of positive statements during the meeting.&RQ1 \\ 
		H3.4$_{0}$ & The negative mood after the meeting does not depend on the amount of negative statements during the meeting.&RQ1 \\
		\midrule
		H4$_0$& \textbf{The perceived probability for conflicts does not depend on the polarity of the statements during the meeting.} &RQ2\\
		H4.1$_{0}$ & The perceived likelihood for social conflicts does not depend on the amount of positive statements during the meeting.&RQ2 \\ 
		H4.2$_{0}$ & The perceived likelihood for social conflicts does not depend on the amount of negative statements during the meeting.&RQ2 \\
		H4.3$_{0}$ & The perceived likelihood for task-related conflicts does not depend on the amount of positive statements during the meeting.&RQ2 \\ 
		H4.4$_{0}$ & The perceived likelihood for task-related conflicts does not depend on the amount of negative statements during the meeting.&RQ2\\
		\midrule
		H5$_0$ & \textbf{There is no difference in the amount of positive statements in the first quarter of the meeting compared to the last quarter.}& RQ3\\
		\midrule
		H6$_0$ & \textbf{There is no difference in the amount of negative statements in the first quarter of the meeting compared to the last quarter.}&RQ3\\
		\midrule 
		H7$_0$ & \textbf{The polarity of the first quarter of the statements during the meeting does not depend on the mood before the meeting.}&RQ3\\
		H7.1$_{0}$ & The amount of positive statements in the first quarter of the meeting does not depend on the positive mood before the meeting.&RQ3 \\ 
		H7.2$_{0}$ & The amount of positive statements in the first quarter of the meeting does not depend on the negative mood before the meeting. &RQ3\\ 
		H7.3$_{0}$ & The amount of negative statements in the first quarter of the meeting does not depend on the positive mood before the meeting. &RQ3\\ 
		H7.4$_{0}$ & The amount of negative statements in the first quarter of the meeting does not depend on the negative mood before the meeting.&RQ3 \\ 
		\midrule
		H8$_0$ & \textbf{The mood after the meeting does not depend on the polarity of the last quarter of statements during the meeting.}&RQ3\\
		H8.1$_{0}$ & The positive mood after the meeting does not depend on the amount of positive statements in the last quarter of  the meeting. &RQ3\\ 
		H8.2$_{0}$ & The positive mood after the meeting does not depend on the amount of negative statements in the last quarter of the meeting.&RQ3 \\
		H8.3$_{0}$ & The negative mood after the meeting does not depend on the amount of positive statements in the last quarter of the meeting.&RQ3 \\ 
		H8.4$_{0}$ & The negative mood after the meeting does not depend on the amount of negative statements in the last quarter of the meeting.&RQ3 \\
		\midrule
		\multicolumn{3}{l}{\textbf{Relationship between the hypotheses and the variables}}\\
		\midrule
		H1 & Positive statements ($m_{pos}$), negative statements ($m_{neg}$)&\\
		H2 & Positive statements ($m_{pos}$), negative statements ($m_{neg}$), positive group affect ($PA(T, t_1)$), negative group affect ($NA(T, t_1)$)& \\
		H3 & Positive statements ($m_{pos}$), negative statements ($m_{neg}$), positive group affect ($PA(T, t_2)$), negative group affect ($NA(T, t_2)$)& \\
		H4 & Positive statements ($m_{pos}$), negative statements ($m_{neg}$), social conflicts ($SC(T)$), task-related conflicts ($TC(T)$) &\\
		H5 & Positive statements ($m_{pos25}, m_{pos75}$)&\\
		H6 & Negative statements ($m_{neg25}, m_{neg75}$)&\\
		H7 & Positive statements ($m_{pos25}$), negative statements ($m_{neg25}$), positive group affect ($PA(T, t_1)$), negative group affect ($NA(T, t_1)$)& \\
		H8 & Positive statements ($m_{pos75}$), negative statements ($m_{neg75}$), positive group affect ($PA(T, t_2)$), negative group affect ($NA(T, t_2)$)& \\ 
		\bottomrule 
	\end{tabularx} 
\end{table*}

\begin{figure*}[htbp]	
	\centering
	\includegraphics[width=0.7\linewidth]{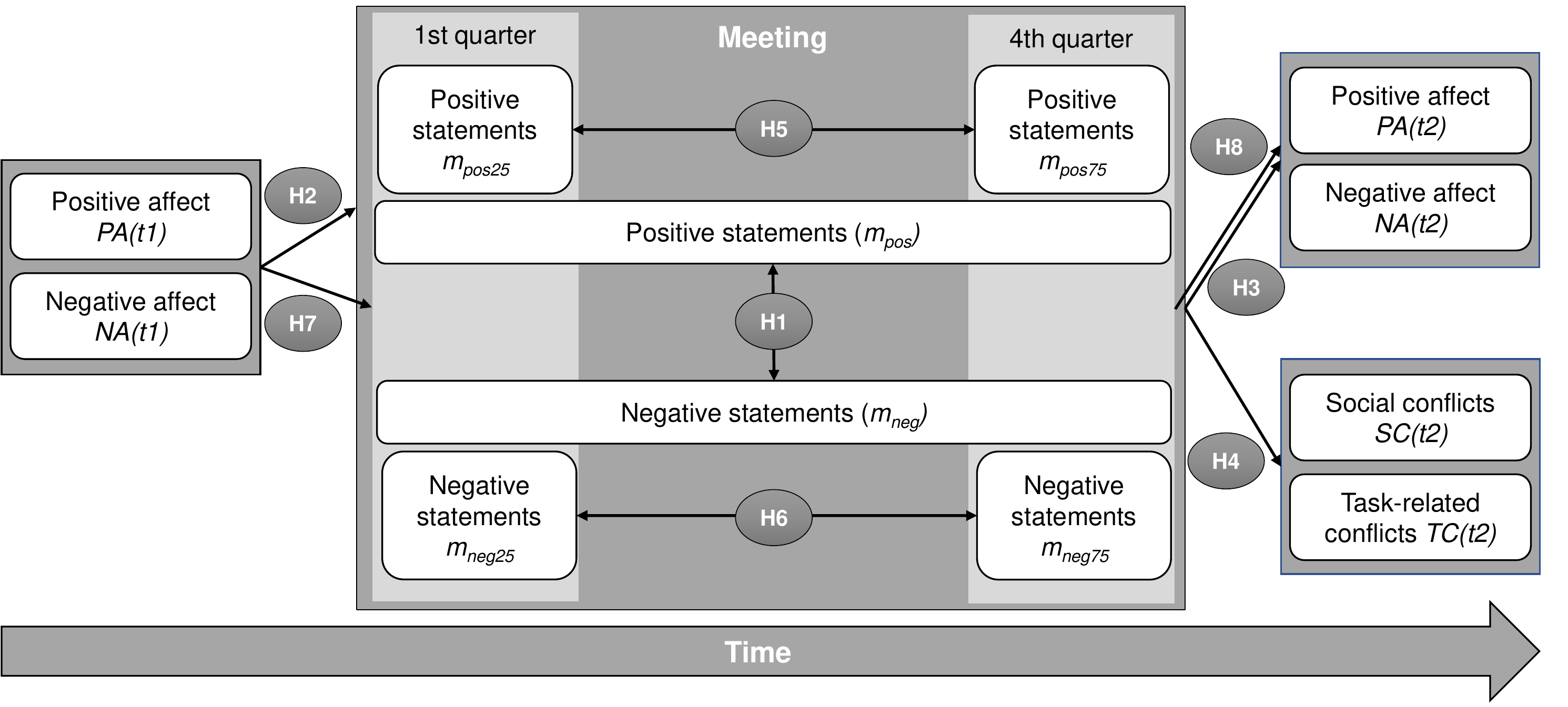}
	\caption{Overview of the variables and the hypotheses}
	\label{fig:overview-hypotheses}
\end{figure*}

\begin{table*}[b]
	\small
	\caption{Overview of variables used for our analysis}
	\label{tab:variables}
	\begin{tabularx}{\textwidth}{lXlll}
		\hline
		Variable & Name & Scale & Value Range & Description \\ \hline
		$m_i$ & Meeting $i$ &  & $i \in \lbrace 1..21 \rbrace$ &  \\
		$S_m$ & Set of statements in a meeting $m$ &  &  &  \\
		$pol(s)$ & Polarity of a statement $s$ & nominal & \{-1; 0; 1\} & \\
		$S^+_m$ & Set of positive statements of a meeting $m$ &  &  & Subset of $S_m$ where $pol(s) = 1$ \\
		$S^-_m$ & Set of negative statements of a meeting $m$ &  &  & Subset of $S_m$ where $pol(s) = -1$ \\
		$m_{pos}$ & Amount of positive statements in a meeting $m$ & interval & 0..1 & $\nicefrac{|S^+_m|}{|S_m|}$ \\
		$m_{neg}$ & Amount of negative statements in a meeting $m$ & interval & 0..1 & $\nicefrac{|S^-_m|}{|S_m|}$ \\
		$m_{pos25}$ & Amount of positive statements in the first quarter of $m$ & interval & 0..1 & \\
		$m_{neg25}$ & Amount of negative statements in the first quarter of $m$ & interval & 0..1 &  \\
		$m_{pos75}$ & Amount of positive statements in the last quarter of $m$ & interval & 0..1 & \\
		$m_{neg75}$ & Amount of negative statements in the last quarter of  $m$ & interval & 0..1 &  \\
		\hline
		$p_i$ & Member of a team &  & $i \in \lbrace 1..5 \rbrace$ &  \\
		$T_i$ & Team &  & $i \in \lbrace 1..21\rbrace$ &  \\
		$t$ & Time & & $t_1$, $t_2$ & $t_1$ = before the meeting, $t_2$ = after the meeting \\
		$PA(p, t)$ & Positive affect of a person $p$ at time $t$ & ordinal & 1..5 & Median of the ten items for positive affect \\
		$	NA(p, t)$ & Negative affect of a person $p$ at time $t$ & ordinal & 1..5 & Median of the ten items for negative affect \\ 
		$PA(T, t)$ & Positive affect of a team $T$ at time $t$ & ordinal & 1..5 & median $\lbrace PA(p, t): p $ is a member of $T\rbrace$ \\
		$NA(T, t)$ & Negative affect of a team $T$ at time $t$ & ordinal & 1..5 & median$\lbrace NA(p, t): p$ is a member of $T\rbrace$ \\
		\hline
		$SC(p)$ & Social conflicts rated by a person $p$ & ordinal & 1..6 & Median of the four items for social conflicts \\
		$TC(p)$ & Task-related conflicts rated by a person $p$& ordinal & 1..6 & Median of the four items for task-related conflicts \\ 
		$SC(T)$ & Social conflicts of a team $T$ & ordinal & 1..6 & median $\lbrace SC(p): p$ is a member of $T\rbrace$ \\
		$TC(T)$ &  Task-related conflicts of a team $T$ & ordinal & 1..6 & median $\lbrace TC(p): p $ is a member of $T\rbrace$ \\
		\hline
	\end{tabularx}
\end{table*}

\subsection{Data Analysis}\label{sec:data-analysis}
We followed a three-step data analysis approach consisting of sentiment analysis on the meeting transcripts, a quantitative analysis on team level and a descriptive analysis on person level. In particular, we have eight main hypotheses (see Table~\ref{tab:HypothesisOverview}) which we explain in detail below with their relationships to variables used. Figure~\ref{fig:overview-hypotheses} provides an overview of these relationships for better understanding. In Table~\ref{tab:variables}, we also describe all variables which we calculated in the respective way to investigate the hypotheses in our analysis.

\subsubsection{Sentiment Analysis on Meetings}
We applied sentiment analysis on the meeting transcripts to get an overview of the polarity of the statements made during the meetings. As far as we know, there is only one software engineering sentiment analysis tool for the German language: the SEnti-Analyzer by Herrmann and Klünder~\cite{herrmann2021textual}, originating from a tool introduced by Klünder et al.~\cite{klunder2020identifying}.  

The \textit{SEnti-Analyzer} is meant to apply sentiment analysis to software project meetings. The tool processes meeting audio data and automatically transcribes audio files \cite{herrmann2021textual}. Afterwards, a lexicon-based algorithm classifies the statements according to their polarity. For the classification, the sentiment analysis tool by Klünder et al.~\cite{klunder2020identifying} is applied to the transcript. This classifier is based on text messages from an industrial software project and, hence, are related to software engineering. We adjusted the tool to be applicable in our context. For example, we removed the audio processing. This way, we could use the transcripts as input for the sentiment analysis. 


In the end of this analysis step, we retrieved three percentage values per meeting: the amount of (1) positive, (2) negative, and (3) neutral statements in the meeting. As we assume that neutral statements do not lead to a shift in the mood of a team, we only consider the amount of positive and negative statements of the meeting, denoted as $m_{pos}$ and $m_{neg}$, for further analysis. 


To compare the use of positive and negative statements during the meeting, we test the hypothesis H1 presented in Table~\ref{tab:HypothesisOverview} at a significance level of $p = 0.05$. First, we test the data for normal distribution using the Shapiro-Wilk test~\cite{shapiro1965analysis}. In case of normal distribution, we use the repeated-measures t-test \cite{Student1908}. Otherwise, we use the Wilcoxon signed-rank test~\cite{wilcoxon1992individual}.

\subsubsection{Quantitative Analysis of the Whole Meeting}
In the second step, we analyzed the relation of different aspects on team level using hypothesis testing. For this step, we used the variables $m_{pos}$, $m_{neg}$, $PA(T, t_1)$, $PA(T, t_2)$, $NA(T, t_1)$, $NA(T, t_2)$, $SC(T)$ and $TC(T)$ (cf. Table~\ref{tab:variables}), where $t_1$ is before and $t_2$ is after the meeting. According to Table~\ref{tab:variables}, most these variables are ordinal-scaled, as we calculated the median of a Likert scale. Only the amount of positive and negative statements during the meeting, $m_{pos}$ and $m_{neg}$, are interval-scaled, each representing a percentage value ranging from 0\% to 100\%. These scales influence the choice of statistical tests. 

In order to analyze the relation between the given variables, we tested the hypotheses H2 to H4 presented in Table~\ref{tab:HypothesisOverview} using Spearman's $\rho$. We analyzed the hypotheses at a significance level of $p \leq 0.05$. However, as we tested four hypotheses for each of the hypotheses H2, H3, and H4, we applied the Bonferroni correction \cite{haynes2013} leading to an adjusted significance level of $p_{corr} = p/4 = 0.05/4 = 0.0125$.

\subsubsection{Quantitative Analysis of Meeting Start and End}
Based on two assumptions, we wanted to look into details of the start and the end of the meeting:\\
~\\
\textbf{Assumption 1:} The mood of a team before a meeting influences the first minutes rather than the whole meeting.\vspace{3mm}\\
\textbf{Assumption 2:} The mood of a team after a meeting is influenced by the last minutes rather than by the whole meeting.
~\\\\
These two assumptions lead to further hypotheses. First, we analyzed differences between the use of positive and negative statements in the beginning and in the end of a meeting, leading to the hypotheses H5 and H6 presented in Table~\ref{tab:HypothesisOverview}. We decided to use the first and the last quarter of statements. This selection may impact the results (using the first and the last third would lead to other results). We discuss the consequences following from this selection in Section~\ref{sec:threats-to-validity}. As described in Section~\ref{sec:data-analysis}, we checked the data for normal distribution using the Shapiro-Wilk test~\cite{shapiro1965analysis}. In case of normal distribution, we used the repeated-measures t-test \cite{Student1908} to test the hypothesis. Otherwise, we used the Wilcoxon signed-rank test~\cite{wilcoxon1992individual}. Second, we tested the hypotheses H7 and H8 presented in Table~\ref{tab:HypothesisOverview}. These two hypotheses ask for a relationship between the mood and the polarity of statements by just considering the first and the last quarter of a meeting. Again, we tested the hypotheses using Spearman's $\rho$ at a corrected significance level of $p_{corr} = 0.0125$.


\subsection{Validity Procedures}\label{sec:validity-procedures}
We implemented some validity procedures to reduce and mitigate some threats to validity.~\\

While the first author performed all the statistical analysis and the hypotheses testing, each step was carefully reviewed by the second author of the paper. We further applied rigorous data cleaning and selection procedures, being rather pessimistic than optimistic, in order to avoid using incomplete data. The data cleaning and selection process was done manually by one author and again reviewed by the other one. 

When aggregating the results, we used the median instead of the mean, as the data emerged from a Likert scale. Calculating the mean would have led to more detailed results, which might have been interesting for this analysis. However, from a mathematical viewpoint, we went for the median in order to avoid over-interpreting our results as the median is more robust against outliers. Besides, the mean is not defined on ordinal scales that are not interval scaled.

We also used a significance level of $p = 0.05$, which is a common procedure in software engineering research. As we tested four hypotheses pointing to one main hypothesis, we applied the Bonferroni correction, leading to an adjusted $p$-value of $p_{corr} = 0.0125$.~\\

These implemented validity procedures mitigated some threats to validity. However, there are still some threats we could not influence. These threats to validity will be discussed in Section~\ref{sec:threats-to-validity}. 

\section{Results}\label{sec:results}
We performed all steps of the data analysis as described in Section~\ref{sec:data-analysis}. In the following, we describe the results. 

\subsection{Sentiment Analysis on Meetings}
\begin{figure}[h]
	\centering
	\includegraphics[width=1\linewidth]{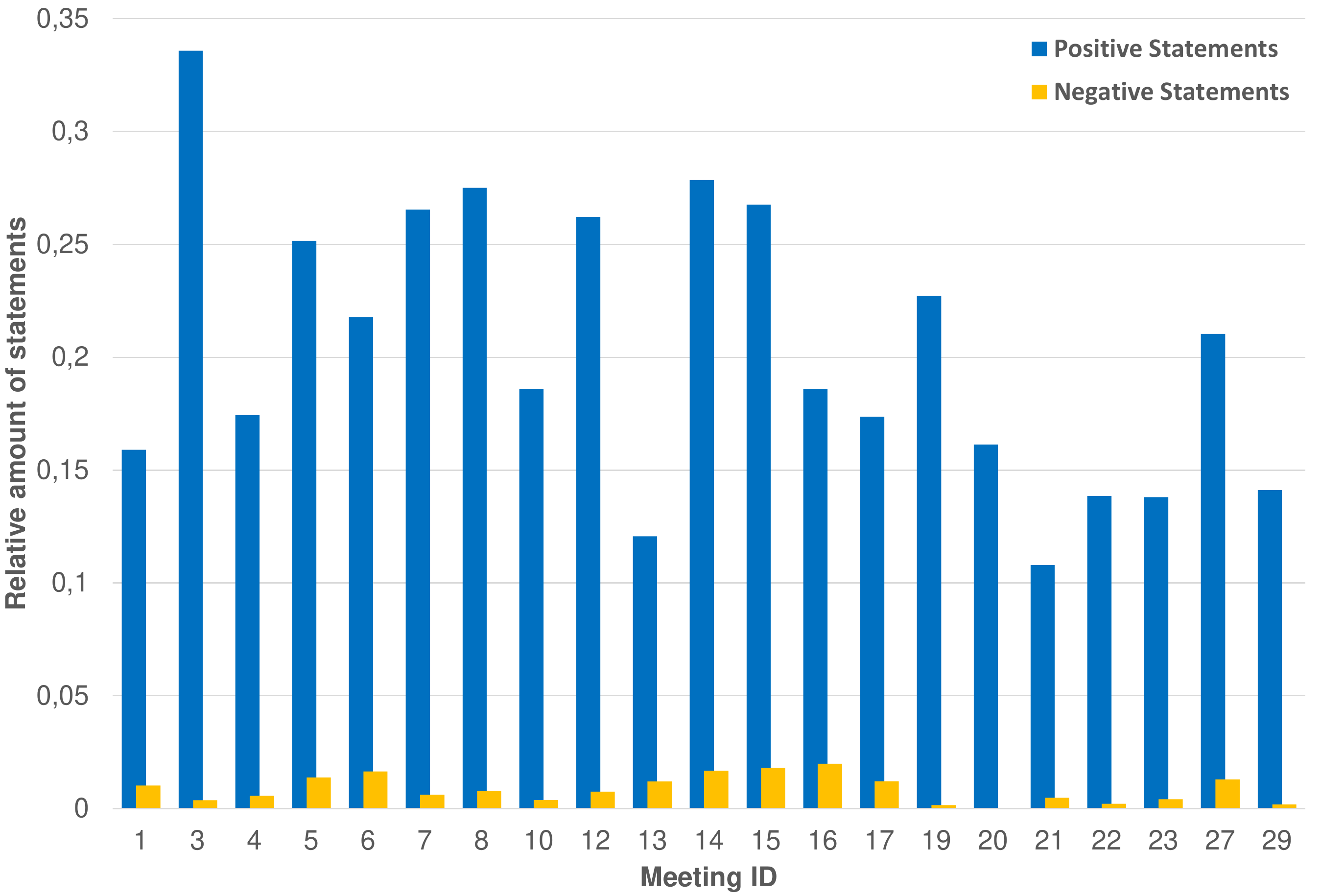}
	\caption{Positive and negative statements per meeting}
	\label{fig:polarity-meetings}
\end{figure}

The students made on average 742 statements during the meetings (min: $162$, max: $1858$, SD: $394$). The amount of positive and negative statements for the meetings is visualized in Figure~\ref{fig:polarity-meetings}. On average, 20.37\% of the statements were positive, with a minimum of 10.79\% and a maximum of 33.58\%. Compared to the positive statements, very few statements were negative: The average amount of negative statements is 0.87\%, with a minimum of 0\% and a maximum of 1.99\%.

The Shapiro-Wilk test showed that the amount of positive and negative statements are both normally distributed (positive: $W=0.95536, p = 0.42808$; negative: $W = 0.93946, p = 0.2127$). Therefore, we used the repeated-measures t-test to test hypothesis H1 (see Table~\ref{tab:HypothesisOverview}). The t-test showed a significant result ($t = -14.68, p < 0.00001$). Therefore, we can reject H1$_0$ and conclude that there is a significant difference in the amount of positive and negative statements in the meeting. Comparing the mean values of positive and negative statements, we see that positive statements are used significantly more than negative statements. 

\vspace{0.7em}
\noindent
\doublebox{
	\begin{minipage}{0.94\columnwidth}
		\textbf{Finding 1:} Intuitively, most student software project teams communicated positively or neutrally, with a maximum amount of 2\% of negative statements in the meeting. \\
		\textbf{Finding 2:} The amount of positive statements is significantly higher than the amount of negative statements.
	\end{minipage}
}

\subsection{Quantitative Analysis on Team Level}
We calculated Spearman's $\rho$  to investigate the hypotheses H2, H3, and H4 using the variables described in Table~\ref{tab:HypothesisOverview}. The results of the tests are summarized in Table~\ref{tab:HypothesisResults}. 

\begin{table}[!bp]
	\small
	\caption{Results of the hypotheses tests}
	\label{tab:HypothesisResults}
	\begin{tabularx}{\columnwidth}{llX}
		\toprule 
		Hyp. & Results & Interpretation\\
		\midrule
		H2&&no statement\\
		H2.1&$r = 0.4664; p (2-$tailed$) = 0.0331$&reject H2.1$_0$\\
		H2.2&$r = -0.0026; p (2-$tailed$) = 0.9910$&no statement\\
		H2.3&$r = -0.084; p (2-$tailed$) = 0.7182$&no statement\\
		H2.4&$r = 0.138; p (2-$tailed$) = 0.5544$&no statement\\
		\midrule
		H3&&no statement\\
		H3.1&$r = 0.2455; p (2-$tailed$) = 0.2835$&no statement\\
		H3.2&$r = -0.2820; p (2-$tailed$) = 0.2156$&no statement\\
		H3.3&$r = -0.2954; p (2-$tailed$) = 0.1936$&no statement\\
		H3.4&$r = -0.2585; p (2-$tailed$) = 0.2579$&no statement\\
		\midrule
		H4&&no statement\\
		H4.1&$r = -0.348; p (2-$tailed$) = 0.1216$&no statement\\
		H4.2&$r = -0.439; p (2-$tailed$) = 0.0464$&reject H4.2$_0$\\
		H4.3&$r = -0.5161; p (2-$tailed$) = 0.0166$&reject H4.3$_0$\\
		H4.4&$r = -0.532; p (2-$tailed$) = 0.01299$&reject H4.4$_0$\\
		\bottomrule 
	\end{tabularx} 
\end{table}

We find no evidence for an influence of the the mood before the meeting on the polarity of statements during the meeting. Hypothesis H2 cannot be rejected, as the only $p$-value that would be considered significant (H2.1 with a $p$-value of $0.0331$) is still above the adjusted $p_{corr}$-value of $0.0125$. Therefore, H2.1 can be rejected (indicating a relation between the amount of positive statements and the positive mood of team before the meeting), but we cannot generalize this insight to make a statement about H2. 

\vspace{0.7em}
\noindent
\doublebox{
	\begin{minipage}{0.94\columnwidth}
		\textbf{Finding 3:} We find no evidence for an influence of the mood before the meeting and the polarity of statements during the meeting.\\
		\textbf{Finding 4:} However, the amount of positive statements depends on the positive mood before the meeting, showing a positive tendency: The higher the positive affect, the more positive statements.  
	\end{minipage}
}
\vspace{0.7em}

\noindent
Regarding the influence of the polarity of statements during the meeting and the mood afterwards, we again find no evidence. As none of the statistical tests concerning H3, namely the tests for H3.1 to H3.4, were significant (neither at the ``normal'' nor at the adjusted $p$-value), we can neither reject H3 nor one of the other hypotheses H3.1 to H3.4.  

\vspace{0.7em}
\noindent
\doublebox{
	\begin{minipage}{0.94\columnwidth}
		\textbf{Finding 5:} We find no evidence for an influence of the polarity of statements during the meeting on the mood of a team after the meeting. 
	\end{minipage}
}
\vspace{0.7em}

\noindent For H4, i.e. the relation between the probability for conflicts and the polarity of the statements, we again find no significant test with a $p$-value that is smaller than $p_{corr}$. Therefore, we cannot reject H4$_0$. However, three of the four hypotheses H4.1 to H4.4 are considered significant at a significance level of $p = 0.05$: We can reject H4.2$_0$, H4.3$_0$, and H4.4$_0$, leading to the following findings:

\vspace{0.7em}
\noindent
\doublebox{
	\begin{minipage}{0.94\columnwidth}
		\textbf{Finding 6:} The perceived likelihood for social conflicts depends on the amount of negative statements during the meeting.\\
		\textbf{Finding 7:} The perceived likelihood for task-related conflicts depends on the amount of positive statements during the meeting.\\
		\textbf{Finding 8:} The perceived likelihood for task-related conflicts depends on the amount of negative statements during the meeting.
	\end{minipage}
}
\vspace{0.7em}

\noindent
As all calculated $r$-values for H4.2 to H4.4 are negative, we have a negative influence. That is, the perceived likelihood for task-related conflicts decreases with an increasing amount of positive or negative statements, and the likelihood for social conflicts increases with a decreasing amount of negative statements, and vice versa. 

\subsection{Quantitative Analysis of Meeting~Start\&End}
In order to test the hypotheses H5 and H6 (see Table~\ref{tab:HypothesisOverview}) on the differences between meeting start and end, we first analyzed the data with respect to normal distribution. The Shapiro-Wilk test showed that two variables for the positive statements are normally distributed ($m_{pos25}: W = 0.9353, p = 0.1758; m_{pos75} = 0.9557, p = 0.4335$), whereas the other two are not ($m_{neg25}: W=0.80635; p = 0.0008; m_{neg75}: W=0.8554, p = 0.0053$). Therefore, we tested H5 using the repeated-measures t-test, and we used the Wilcoxon signed-rank test for H6.\\

The t-test for H5 was significant ($t = -2.819, p = 0.0106$). Therefore, we reject H5$_0$ and assume that there is a difference in the amount of positive statements comparing the start and the end of a meeting. The mean amount of positive statements in the first quarter of the meeting is 22.43\% (min: 14.56\%, max: 36.76\%), and the mean in the last quarter of the meeting is 19.44\% (min: 9.6\%, max: 33.58\%). These values raise the impression that the amount of positive statements in the beginning is significantly higher than in the end, but given the small difference in the mean value, this finding should be handled with care.


\vspace{0.7em}
\noindent
\doublebox{
	\begin{minipage}{0.94\columnwidth}
		\textbf{Finding 9:} There is a significant difference between the amount of positive statements comparing the first and the last quarter of the meeting. \\
		\textbf{Finding 10:} On average, in the first quarter, 22.43\% of the statements are positive, which is slightly more than the average for the last quarter of 19.44\%.
	\end{minipage}
}
\vspace{0.7em}

\noindent
To compare the amount of negative statements in the first and the last quarter of the meeting, we used the Wilcoxon signed-rank test. The test did not result in a significant result ($z = -0.9087, p = 0.3628$). Thus, we cannot reject H6$_0$. 

\vspace{0.7em}
\noindent
\doublebox{
	\begin{minipage}{0.94\columnwidth}
		\textbf{Finding 11:} We find no difference between the amount of negative statements when comparing the first and the last quarter of the meeting.
	\end{minipage}
}
\vspace{0.7em}

\noindent
In a next step, we tested the hypotheses H7 and H8 presented in Table~\ref{tab:HypothesisOverview}. The results of Spearman's $\rho$ are summarized in Table~\ref{tab:HypothesisResults-II}. 

\begin{table}[!htbp]
	\small
	\caption{Results of the hypotheses tests}
	\label{tab:HypothesisResults-II}
	\begin{tabularx}{\columnwidth}{llX}
		\toprule 
		Hyp. & Results & Interpretation\\
		\midrule
		H7&&no statement\\
		H7.1&$r = 0.4751; p (2-$tailed$) = 0.0295$&reject H7.1$_0$\\
		H7.2&$r = -0.072; p (2-$tailed$) = 0.7569$&no statement\\
		H7.3&$r = -0.076; p (2-$tailed$) = 0.7418$&no statement\\
		H7.4&$r = 0.5043; p (2-$tailed$) = 0.0197$&reject H7.4$_0$\\
		\midrule
		H8&&no statement\\
		H8.1&$r = 0.1879; p (2-$tailed$) = 0.4147$&no statement\\
		H8.2&$r = -0.1284; p (2-$tailed$) = 0.5792$&no statement\\
		H8.3&$r = -0.1847; p (2-$tailed$) = 0.423$&no statement\\
		H8.4&$r = -0.2633; p (2-$tailed$) = 0.2488$&no statement\\
		\bottomrule 
	\end{tabularx} 
\end{table} 

We find significant results for H7.1 and H7.4. Therefore, we can reject H7.1$_0$ and H7.4$_0$ and conclude that there is an influence of the positive mood of a team before the meeting and the amount of positive statements in the first quarter (and respectively for negative statements). However, as both $p$-values are above the adjusted $p$-value of $0.0125$, we cannot reject H7. 

\vspace{0.7em}
\noindent
\doublebox{
	\begin{minipage}{0.94\columnwidth}
		\textbf{Finding 12:} The positive mood of a team before the meeting influences the amount of positive statements in the first quarter of the meeting.\\
		\textbf{Finding 13:} The same is correct for negative statements: The negative mood of a team influences the amount of negative statements in the first quarter.
	\end{minipage}
}
\vspace{0.7em}

\noindent
Regarding H8, we find no significant results. Thus, we can neither reject H8, nor one of the hypotheses H8.1 to H8.4.


\vspace{0.7em}
\noindent
\doublebox{
	\begin{minipage}{0.94\columnwidth}
		\textbf{Finding 14:} We find no evidence for an influence of the statements made in the last quarter of the meeting and the mood afterwards.
	\end{minipage}
}

\section{Discussion}\label{sec:discussion}
Based on our results, we conclude this paper by answering the research questions, discussing our results, and
presenting the threats to validity.

\subsection{Answering the Research Questions}
According to the results emerging from 21 student software projects, we can answer the three research questions posed in Section~\ref{sec:research-goal-and-research-questions} as follows:\\

\noindent
\textbf{Answer to RQ1:} Surprisingly, we only find evidence for a very few relations between the three parameters (1) mood before the meeting, (2) polarity of statements during the meeting, and (3) mood after the meeting. According to our results, positive mood before the meeting and the amount of positive statements during the meeting are related to each other (and given the timeframe, positive mood influences the amount of positive statements). We do not find any additional significant influences between these three parameters.\\

\noindent
\textbf{Answer to RQ2:} Regarding the polarity of statements made during the meeting and the likelihood of conflicts afterwards, our results show a relation between these parameters. There is a significant influence of both positive and negative statements on task-related conflicts, as well as an influence of positive statements on social conflicts. All these relations are negative, that is the more positive/negative statements, the less conflicts and vice versa.\\

\noindent
\textbf{Answer to RQ3:} Only considering the first and the last quarter of the meeting, we find an influence of the positive mood of a team before the meeting on the amount of positive statements, and an influence of the negative mood of a team on the amount of negative statements. Both influences are positive: The higher the positive/negative affect, the higher the amount of positive/negative statements. However, we do not find evidence for an influence of the polarity of the statements on the mood after the meeting. 

\subsection{Interpretation}
Our results allow three noteworthy observations:

\begin{itemize}
	\item Students intuitively behave adequately in meetings, making way more positive than negative statements during the meetings. 
	\item Starting a meeting with a high positive mood can smooth both the meeting start as well as the meeting as a whole.
	\item The polarity of statements made during the meeting has no measurable influence on the mood afterwards. 
\end{itemize}

While the two first observations are self-explanatory and intuitive, the last one is not that obvious. According to the results of our study, a positive mood has an influence on the amount of positive statements both at meeting start and throughout the meeting. However, as mood is a long-term affective state, it is difficult to change it right before a meeting. In future research, it would be interesting to conduct the same study focusing on short-term emotional states, i.e., emotions instead of affect, as they can be ``changed'' more easily (e.g., by bringing sweets to the meeting).

Another example for interesting findings with a significant influence are Findings 7 and 8: Both findings show an influence of positive respectively negative statements on the perceived likelihood of task-related conflicts. Looking at the results of the statistical tests, we observe an influence in the same direction (both $r$-values are below 0). Therefore, we have an influence of both negative and positive statements on task-related conflicts showing the desired tendency. That is, the more positive (or negative) statements, the lower the perceived likelihood for task-related conflicts. Consequently, one can conclude that it does not matter whether the statements are positive or negative, but they should not be neutral to lower the risk for task-related conflicts. One possible explanation is that both positive and negative statements can be used to point to a task-related conflict (e.g., given by unclear tasks or misunderstood requirements). It is less important how the meeting participants point to these issues, as long as they are mentioned at all.

Besides the significant results we found in our data, it is interesting which results we did not see:
Why is there no influence of the meeting on the mood afterwards? None of the hypotheses H3.x and H8.x are significant. There are two possible explanations: First, it is possible that there is an influence that we simply do not see in our data. This can be due to a too small $n$, the nature of student projects, or the study itself. But, there is another possible explanation emerging when we look at the definition of mood and emotion.
Psychologists distinguish between two definitions: \textit{emotion} and \textit{mood}.
While an emotion is a short-term and rather intense experience~\cite{sander2009oxford}, mood differs from this in duration and intensity. According to Ekkekakis and Russel~\cite{ekkekakis2013measurement}, mood is an affective state more persistent than an emotional state, and hence more stable. This difference between short-term emotional state and long-term affective state might explain the results of our analysis.

In our analysis, we used PANAS (positive and negative affect schedule) that is a measure for long-term affective states. Looking at the definition of mood, it is a long-term affective state~\cite{russell1999core}. Therefore, it is possible (and not unlikely) that the short-term emotional state (emotions) after a meeting changes, if there were too many negative statements in the meeting, but this does not necessarily lead to an increase in the negative long-term affective state (mood). Consequently, it is likely that ``bad behavior'' in meetings (in terms of negative statements) can affect the short-term emotional state after the meeting, but has no (or only marginal) influence on the long-term affective state. The meeting itself is just a short point in time compared to the longer time frame that has an influence on the long-term affective state (PANAS often asks for a rating over a period of one week~\cite{schneider2015media}). Therefore, the meeting may change the emotions of the participants afterwards, but has not a measurable influence on the mood.

Therefore, looking at future meetings, despite the fact that we do not observe an influence of the statements in the meeting on the mood afterwards, one must not assume that there is no influence. We considered the influence on the long-term affective state, but it is not unlikely that short-term emotional states are influenced by the behavior in meetings. However, it is worth a thought to pay more attention to effective meetings than to an adequate behavior (which should be natural).

Considering other non-significant results, it is kind of soothing that negative mood before the meeting has no influence on the polarity of the statements made during the meeting. There are again two possible explanations: First, it is possible that the variance in the negative mood is rather small, so that all participants were in a rather ``good mood'' (i.e. having a higher positive than negative affective state). Second, it is possible that the student meeting participants were sufficiently professional that their negative affective state did not influence the whole meeting. As we find a significant relation between the negative mood before a meeting and the amount of negative statements in the first quarter of the meeting, it might be that the students are not able to completely compensate a ``bad mood'', but are able to refocus  after a couple of minutes, forgetting the reasons for their negative affective state. 

\subsection{Future Work}
According to our results, there is no relation between the meeting and the mood. However, as discussed before, it is likely that the short-term emotional states are affected by the polarity of the statements in the meetings, and, thus, by the meeting itself. Therefore, future studies focusing on this short-term effect are necessary to analyze this assumption in more detail. For example, a similar study using the wheel of emotion~\cite{plutchik1991emotions} to measure the short-term emotional state would be interesting and helps investigate the relation between meetings and mood in more detail.

In addition, a more diverse sentiment analysis not only focusing on polarities would also be interesting to draw conclusions on the long-term affective state after a meeting. That is, increasing the granularity of the polarities of the statements (considering joyful, friendly, angry, or other types of statements instead of positive, negative, or neutral) would lead to more fine-grained results and can uncover insights that are impossible with the analysis presented in this paper.    

Besides, given the increasing amount of online meetings, it would also be interesting to conduct a similar study in remote or hybrid settings, assuming that interaction patterns differ between different types of meetings. In line with this thought, it would also be helpful to replicate such studies with different teams in different contexts (experience, duration of collaboration, and similar).

Based on these more detailed analyses, it would be possible to provide concrete recommendations for future meetings to increase the chance for a satisfied software project team. These recommendations may include information on wether it is meaningful to assess a team's mood before a meeting (and how to do so), and what to do in case of negative mood before a meeting (or in case of too good mood, e.g., due to holidays starting soon).

However, afterwards, the usefulness for industry needs to be evaluated in real-world settings. That is: Do practitioners need support to identify negative statements in order to avoid negative effects (that, indeed, do not seem to be related to the statements)? Or are they aware of the relevance of their meeting behaviour? Consequently, an interview study in industry on the difficulties in software projects caused by or related to social aspects can point to future research directions. 

In addition, there are several other open questions regarding meetings and social aspects in software projects: How are decisions during a meeting influenced by the overall team mood? That is, does a team decide differently in dependence of the mood? More concretely, it would be interesting to analyze if a negative mood lead to lower quality decisions. 

To the best of our knowledge, there is, so far, no study investigating pain points in industry that are related to social aspects. Research has often proven relationships between social aspects and project-related aspects (cf. \cite{klunder2018helping,kortum2017don}), but it remains unclear what research needs to provide in order to be applicable and useful for industry.

\subsection{Threats to Validity}\label{sec:threats-to-validity}
Despite the implemented validity procedures described in Section~\ref{sec:validity-procedures}, the results of our study are subject to some limitations and threats to validity. In this section, we discuss the threats according to the classification by Wohlin et al.~\cite{wohlin2012experimentation}. 

\paragraph{Construct Validity} 
The results of our study are based on a data collection by Schneider et al.~\cite{schneider2015media,schneider2018positive}. The data was collected using paper-pencil-questionnaires and established scales from psychology. Given the language used in psychology, it is possible that some students did not know how to report, e.g., on their long-term affective state compared to their short-term emotional state. Such misunderstandings, misinterpretations, or misconceptions would have influenced the data collected. However, to reduce the influence of single misunderstandings, we aggregated all data on team level. That is, we first calculated the median on personal level (such that one item had an influence of only 10\% on the median), and then we aggregated the median values again to retrieve the mood. These aggregations reduced the influence of single misunderstood ratings.

\paragraph{Internal Validity}
Survey results reflect personal perspectives and subjective opinions. In particular, the used questionnaires asked for psychological aspects such as mood and conflicts. Answering or rating these items is very subjective and depends on the participants. Consequently, the results cannot be considered objective. 

Before starting the analysis, we carefully cleaned the data and defined the data set to be used for the analysis. That is, we removed incomplete data points as described in Section~\ref{sec:data-selection-and-cleaning}. The data analysis followed a stringent procedure performed by one researcher and reviewed by another one. 

Nevertheless, despite the data cleaning procedures, the threat of the age of the data cannot be mitigated. Our analysis is based on data that was collected almost a decade ago. Although there still are face-to-face meetings, a not negligible amount of meetings takes place in an online setting. We assume that some of our results are also correct for remote settings, but further research is required to allow drawing similar conclusions for today's meetings. 

In addition, only 2\% of the statements in the meeting were negative. Given that team members did not know each other a long time before the meeting, it is possible that the amount of negative statements is higher in teams that already worked together for a long time and, thus, know each other very well. Consequently, future research should take into account different phases of team building. 

\paragraph{Conclusion Validity}
In this paper, we performed statistical analysis and hypotheses testing. Despite the small sample size of 21 teams, we retrieved significant results. We applied the Bonferroni correction if necessary, and also calculated the $p$-value when applying Spearman's $\rho$. This way, we did not only get a correlation coefficient, but also its reliability. 

Nevertheless, our results are only based on a sample size of 21 student software projects. Further research is required to provide deeper insights and to increase the reliability of our results.    

\paragraph{External Validity}
As the data set emerged from a course at Leibniz University Hannover, we had no influence on the participant selection. All of them were computer science students close to their bachelor's degree. Using students as participants always reduces the generalizability of the results, as student projects are not necessarily comparable to real industrial projects. However, a first step to analyze the relation between mood and meeting on a basic level requires comparable data sets to draw conclusions. Therefore, despite the impossibility to generalize the results to industry, we used this respective data set.

\section{Conclusion}\label{sec:conclusion}
Using data from 21 student software projects, we analyzed the relation between mood before a meeting, the polarity of the statements during the meeting, the mood after a meeting, and the perceived likelihood for social and task-related conflicts. To provide more fine-grained results, we also divided the meeting into three parts: first quarter, middle, and last quarter, and analyzed the influence between beginning and end, and the respective mood. 

Our results show that positive and negative mood before a meeting are related to the amount of positive respectively negative statements in the beginning of the meeting. However, neither the meeting as a whole nor the last quarter of a meeting are related to the mood afterwards. Nevertheless, the perceived likelihood for social and task-related conflicts depends on the amount of positive and negative statements during the meeting. 

Future research can strengthen the results by using real (industrial) software projects instead of student projects, and by changing the measured aspects. For example, looking at short-term emotional states (emotions) instead of long-term affective states (mood or affect) might provide deeper insights and can help further improve future meetings.

\balance

\bibliographystyle{ACM-Reference-Format}
\bibliography{references}

\end{document}